\documentclass[aps,prd,twocolumn]{revtex4}
\usepackage{pdfpages}
\usepackage{graphicx}
\usepackage{amsmath}
\usepackage{booktabs}
\usepackage{color}
\usepackage{multirow}

\begin{document}
\title{$\boldsymbol{P}$-wave states $\boldsymbol{T^-_{bb}}$ from diquarks}

\author{Zu-Hang Lin, Chun-Sheng An, and Cheng-Rong Deng${\footnote{crdeng@swu.edu.cn}}$}

\affiliation{School of Physical Science and Technology, Southwest University, Chongqing 400715, China}

\begin{abstract}
We investigate the $P$-wave states $T^-_{bb}$ in the isospin singlet and three
excited modes [excitation occurring in the diquark $[bb]^{s_1}_{c_1}$ ($\rho $-mode),
antidiquark $[\bar{u}\bar{d}]^{s_2}_{c_2}$ ($r$-mode) or between them ($\lambda$-mode)]
from diquarks in a quark model. We analyze the dynamical behaviors of the diquark
$[bb]^{s_1}_{c_1}$, antidiquark $[\bar{u}\bar{d}]^{s_2}_{c_2}$ and their correlations
in the states $T^-_{bb}$ by decomposing the interactions from various sources in the
model. The absolute dominant color-spin configuration, more than $99\%$, in the $\rho$-mode
with $1^1P_1$ is $[bb]^0_{\bar{\mathbf{3}}}[\bar{u}\bar{d}]^0_{\mathbf{3}}$. Its
energy is lower by about $18$ MeV than the threshold $\bar{B}\bar{B}$ so that it can
establish a compact bound state. The chromomagnetic and meson-exchange interactions in the
antidiquark $[\bar{u}\bar{d}]^0_{\mathbf{3}}$ are responsible for its binding mechanism.
Two other excited modes are higher than their respective threshold. The color configuration
$\mathbf{6}\otimes\bar{\mathbf{6}}$ need to be handled discreetly in the tetraquark states.
\end{abstract}
\maketitle

\section{Introduction}
Searching for multiquark hadrons is an extremely significant topic in hadronic
physics because they may contain more abundant low-energy strong interaction
information than ordinary hadrons. Theoretical explorations on the possible
stable doubly heavy tetraquark states can be traced back to the early
1980s~\cite{Ader:1981db,Ballot:1983iv}. Their properties have been studied
extensively in various theoretical frameworks in recent years~\cite{Huang:2023jec,
Chen:2022asf}, in particular since the LHCb Collaboration reported the doubly
charmed state $T^+_{cc}$ with $01^+$ in the $D^0D^0\pi^+$ invariant mass
spectrum~\cite{LHCb:2021vvq,LHCb:2021auc}. The doubly heavy tetraquark states
are usually discussed as a diquark-antidiquark or meson-meson configuration. The
former can establish a compact state while the the latter can produce a relative
loose molecular state~\cite{Huang:2023jec,Chen:2022asf}. The tendency to form
stable bound states is proportional to the mass ratio of heavy quark and light
antiquark. The majority of the existing theoretical investigations mainly
concentrate on the doubly heavy tetraquark ground states, which indicates that
the state $T^-_{bb}$ with $01^+$ can establish a deep bound state though it has
not been observed in experiments~\cite{Huang:2023jec,Chen:2022asf}. In this case,
its $P$-wave excited states are most likely to be stable against the strong
interactions in the low excited states of the doubly heavy tetraquark states.
Recently, the $P$-wave excited states $T^-_{bb}$ were explored using the lattice
QCD potential and Born-Oppenheimer approximation~\cite{Bicudo:2017szl,Hoffmann:2022jdx},
quark models~\cite{Kim:2022mpa,Meng:2021yjr}, and the QCD Laplace sum rule~\cite{Albuquerque:2023rrf}.

The concept of diquarks was first proposed by Gell-Mann in his pioneering work on
quarks~\cite{Gell-Mann:1964ewy}. Subsequently, diquarks are usually regarded
as an elementary constituent to explore the properties of hadrons and hadron-hadron
interactions~\cite{Anselmino:1992vg}. It seems that there exists some phenomenological
evidence of the relevance of the diquarks in hadron physics~\cite{Jaffe:2004ph,Barabanov:2020jvn}.
The sizes of diquarks are often ignored for simplicity in some calculations~\cite{Anselmino:1992vg},
just like the sizes of constituent quarks are neglected. In the constituent quark
models, diquarks are not a pointlike fundamental object but a spatially extended
object with various color-spin-isospin-orbit configurations, which makes the hadron
world more fantastic.

The substructures of the diquarks usually affect the structures and properties of
hadrons and the diquark correlations might be critical to the formation of multiquark
hadrons~\cite{Olsen:2017bmm,Guo:2019twa,Liu:2019zoy}. The doubly heavy tetraquark
states can provide a clear diquark picture so that they are extremely beneficial to
research the substructures, natures and correlations of diquarks. Inspired by the state
$T^+_{cc}$ reported by the LHCb Collaboration, the lattice QCD calculation on
the channel $DD^*$ with $01^+$ indicated that its short-range attraction was
related to the attractive diquark color-spin configuration
$[cc]^0_{\bar{\mathbf{3}}}[\bar{u}\bar{d}]^1_{\mathbf{3}}$~\cite{Lyu:2023xro}.
Similar short-range attraction was also found in the channel $\bar{B}\bar{B}^*$
with $01^+$~\cite{Lyu:2023xro}.

In this work, we prepare to explore the natures and structures of the $P$-wave
excited states $T^-_{bb}$ in the isospin singlet and various excited modes from the
perspective of diquarks. We decode the natures of the diquark $[bb]^{s_1}_{c_1}$
and antidiquark $[\bar{u}\bar{d}]^{s_2}_{c_2}$ and their correlations in the
nonrelativistic quark model. The model can well describe the nature of the state
$T^+_{cc}$~\cite{Deng:2021gnb}. The motivation of this paper is to broaden our visions
on the properties and structures of the excited states $T^-_{bb}$ from the perspective
of the phenomenological model. We anticipate providing some valuable information
for the experimental establishment of the doubly heavy tetraquark states in the future.

This paper is organized as follows. After the Introduction, we give the details
of the quark model in Sec. II. We show the wave functions of the states $T^-_{bb}$
in Sec. III. We present the numerical results and discussions in Sec. IV. We list
a brief summary in the last section.

\section{Quark Model}

At the hadron scale, QCD is highly nonperturbative due to the complicated infrared
behavior of the non-Abelian SU(3) gauge group. The calculations of hadron spectra
and the hadron-hadron interaction directly from QCD are very difficult at present.
A less rigorous approach, the QCD-inspired quark model, is a powerful implement
in obtaining physical insight for these complicated strong interacting
systems. The quark model is formulated under the assumption that the hadrons
are color-singlet nonrelativistic bound states of constituent quarks with
phenomenological effective masses and interactions. One expects the model dynamics
to be governed by QCD. The perturbative effect is the well-known one-gluon-exchange
(OGE) interaction. From the nonrelativistic reduction of the OGE diagram in QCD
for point-like quarks one gets
\begin{eqnarray}
V_{ij}^{\rm oge}={\frac{\alpha_{s}}{4}}\boldsymbol{\lambda}^c_{i}
\cdot\boldsymbol{\lambda}_{j}^c\left({\frac{1}{r_{ij}}}-
{\frac{2\pi\delta(\mathbf{r}_{ij})\boldsymbol{\sigma}_{i}\cdot
\boldsymbol{\sigma}_{j}}{3m_im_j}}\right),
\end{eqnarray}
where $\boldsymbol{\lambda}^c_{i}$ and $\boldsymbol{\sigma}_{i}$ stand for the
color $SU(3)$ Gell-Mann matrices and spin $SU(2)$ Pauli matrices, respectively.
$r_{ij}$ is the distance between the quarks $i$ and $j$ and $m_i$ is the mass of
the $i$-th quark. The Dirac $\delta(\mathbf{r}_{ij})$ function should be regularized
in the form~\cite{Vijande:2004he}
\begin{equation}
\delta(\mathbf{r}_{ij})\rightarrow\frac{1}{4\pi r_{ij}r_0^2(\mu_{ij})}e^{-\frac{r_{ij}}{r_0(\mu_{ij})}},
\end{equation}
where $r_0(\mu_{ij})=\frac{\hat{r}_0}{\mu_{ij}}$, $\mu_{ij}$ is the
reduced mass of two interacting particles $i$ and $j$. The quark-gluon
coupling constant $\alpha_s$ adopts an effective scale-dependent
form,
\begin{equation}
\alpha_s(\mu^2_{ij})=\frac{\alpha_0}{\ln\frac{\mu_{ij}^2}{\Lambda_0^2}},
\end{equation}
$\hat{r}_0$, $\Lambda_0$ and $\alpha_0$ are adjustable parameters
fixed by fitting the ground state meson spectrum.

The constituent quark mass originates from the breaking of the $SU(3)_L\otimes SU(3)_R$
chiral symmetry at some momentum scale~\cite{Manohar:1983md}. The chiral symmetry is
spontaneously broken in the light quark sector while it is explicitly broken in the
heavy quark sector. Once the light constituent quark mass is generated, it has
to interact through Gold-stone bosons $\pi$, $K$ and $\eta$. In addition, the scalar
meson $\sigma$ exchange interaction is involved. The explicit Gold-stone boson
exchange potentials $V_{ij}^{\pi}$, $V_{ij}^{K}$, $V_{ij}^{\eta}$ and $\sigma$-meson
exchange potential $V_{ij}^{\sigma}$ are taken from Ref.~\cite{Vijande:2004he},
\begin{eqnarray}
\begin{aligned}
V_{ij}^{\rm obe}= & V^{\pi}_{ij} \sum_{k=1}^3\boldsymbol{F}_i^k
\boldsymbol{F}_j^k+V^{K}_{ij} \sum_{k=4}^7\boldsymbol{F}_i^k\boldsymbol{F}_j^k \\
+&V^{\eta}_{ij} (\boldsymbol{F}^8_i\boldsymbol{F}^8_j\cos \theta_P
-\sin \theta_P),\\
V^{\chi}_{ij}= &
\frac{g^2_{ch}}{4\pi}\frac{m^3_{\chi}}{12m_im_j}
\frac{\Lambda^{2}_{\chi}}{\Lambda^{2}_{\chi}-m_{\chi}^2}
\mathbf{\sigma}_{i}\cdot
\mathbf{\sigma}_{j}  \\
\times &\left( Y(m_\chi r_{ij})-
\frac{\Lambda^{3}_{\chi}}{m_{\chi}^3}Y(\Lambda_{\chi} r_{ij})
\right),~Y(x)=\frac{e^{-x}}{x} \\
V^{\sigma}_{ij}= &-\frac{g^2_{ch}}{4\pi}
\frac{\Lambda^{2}_{\sigma}m_{\sigma}}{\Lambda^{2}_{\sigma}-m_{\sigma}^2}
\left( Y(m_\sigma r_{ij})-
\frac{\Lambda_{\sigma}}{m_{\sigma}}Y(\Lambda_{\sigma}r_{ij})
\right).  \\
\end{aligned}
\end{eqnarray}
$\boldsymbol{F}_{i}$ are the flavor $SU(3)$ Gell-Mann matrices and $\chi$
represents $\pi$, $K$ and $\eta$. The mass parameters $m_{\chi}$ take
their experimental values, while the cutoff parameters $\Lambda_{\chi}$
and the mixing angles $\theta_{P}$ take the values from~\cite{Vijande:2004he}.
The mass parameter $m_{\sigma}$ can be determined through the partial conservation
of axial vector current relation $m^2_{\sigma}\approx m^2_{\pi}+4m^2_{u,d}$~\cite{Scadron:1982eg}.
The chiral coupling constant $g_{ch}$ can be obtained from the $\pi NN$
coupling constant through
\begin{equation}
\frac{g_{ch}^2}{4\pi}=\left(\frac{3}{5}\right)^2\frac{g_{\pi NN}^2}
{4\pi}\frac{m_{u,d}^2}{m_N^2}.
\end{equation}

In addition to the meson-exchange interactions, the quark model also incorporates
another nonperturbative effect, color confinement, because observed hadrons in
experiments are color singlets. However, it is still impossible to directly
derive color confinement analytically from its QCD Lagrangian so far. From the
perspective of phenomenology, quark confinement potential should only emerge from
a model in which the interaction depends on color charges. In addition, the coupling
between color charges increases with increasing separation. Quark confinement
potential in the quark model can be generally manmade based on the two ingredients.
In this work, the quark confinement potential can be written as
\begin{eqnarray}
\begin{aligned}
V_{ij}^{\rm con}=-a_c\boldsymbol{\lambda}^c_{i}
\cdot\boldsymbol{\lambda}^c_{j}r^2_{ij}.\\
\end{aligned}
\end{eqnarray}

To sum up, the complete Hamiltonian of the quark model for the mesons and $T^-_{bb}$ can be presented as
\begin{eqnarray}
\begin{aligned}
H_n=&\sum_{i=1}^n \left(m_i+\frac{\mathbf{p}_i^2}{2m_i}\right)-T_{c}+\sum_{i<j}^n V_{ij} \\
V_{ij}=&V_{ij}^{\rm oge}+V_{ij}^{\rm con}+V_{ij}^{\rm obe}+V_{ij}^{\sigma},
\end{aligned}
\end{eqnarray}
where $\mathbf{p}_i$ is the momentum of the $i$-th quark and $T_{c}$ is the center-of-mass
kinetic energy. Accurately solving the two-body Schr\"{o}dinger equation, we can obtain a
set of model parameters by fitting the ground state meson spectrum with the Minuit
program~\cite{James:1975dr}, which are presented in Tables~\ref{parameters} and~\ref{meson},
respectively.
\begin{table}[ht]
\caption{Adjustable model parameters, quark mass and $\Lambda_0$
unit in MeV, $a_c$ unit in MeV$\cdot$fm$^{-2}$, $r_0$ unit in MeV$\cdot$fm
and $\alpha_0$ is dimensionless.}\label{parameters}
\tabcolsep=0.158cm
\begin{tabular}{cccccccccccccccccc}
\toprule[0.8pt] \noalign{\smallskip}
Para. & $m_{u,d}$ & $m_{s}$ & $m_c$ & $m_b$ & $a_c$ & $\alpha_0$ & $\Lambda_0$ & $r_0$   \\
\noalign{\smallskip}
Valu. &    280    &  512    & 1602  & 4936  & 40.78 &   4.55     &   9.17      & 35.06   \\
\toprule[0.8pt] \noalign{\smallskip}
\label{Table1}
\end{tabular}
\caption{Ground state meson spectrum, units in MeV. \label{meson}}
\tabcolsep=0.18cm
\begin{tabular}{cccccccccccccccccccccccc}
\toprule[0.8pt] \noalign{\smallskip}
\renewcommand{\arraystretch}{8.0}
State     & $\pi$ & $\rho$ & $\omega$ &  $K$   & $K^*$ & $\phi$ & $D^{\pm}$ \\
\toprule[0.8pt] \noalign{\smallskip}
Theo.     &  142  &  826   &   780    &  492   & 974   &   1112 & 1867      \\
PDG.      &  139  &  775   &   783    &  496   & 896   &   1020 & 1869      \\
\toprule[0.8pt] \noalign{\smallskip}
State     & $D^*$ & $D_s$  & $D_s^*$ & $\eta_c$ & $J/\Psi$ & $B$ & $B^*$  \\
\noalign{\smallskip}
\toprule[0.8pt] \noalign{\smallskip}
Theo.     & 2002  & 1972   &  2140    &  2912  & 3102  &  5251  & 5301      \\
PDG.      & 2007  & 1968   &  2112    &  2980  & 3097  &  5280  & 5325      \\
\toprule[0.8pt]  \noalign{\smallskip}
State     & $B_s$ & $B_s^*$& $B_c$ & $\eta_b$ & $\Upsilon (1S)$  \\
\noalign{\smallskip}
\toprule[0.8pt] \noalign{\smallskip}
Theo.     & 5377  & 5430   &  6261    & 9441  & 9546               \\
PDG.      & 5366  & 5416   &  6275    & 9391  & 9460               \\
\toprule[0.8pt]  \noalign{\smallskip}
\end{tabular}
\end{table}

\section{wave functions}

\begin{figure} [h]
\centering
\resizebox{0.28\textwidth}{!}{\includegraphics{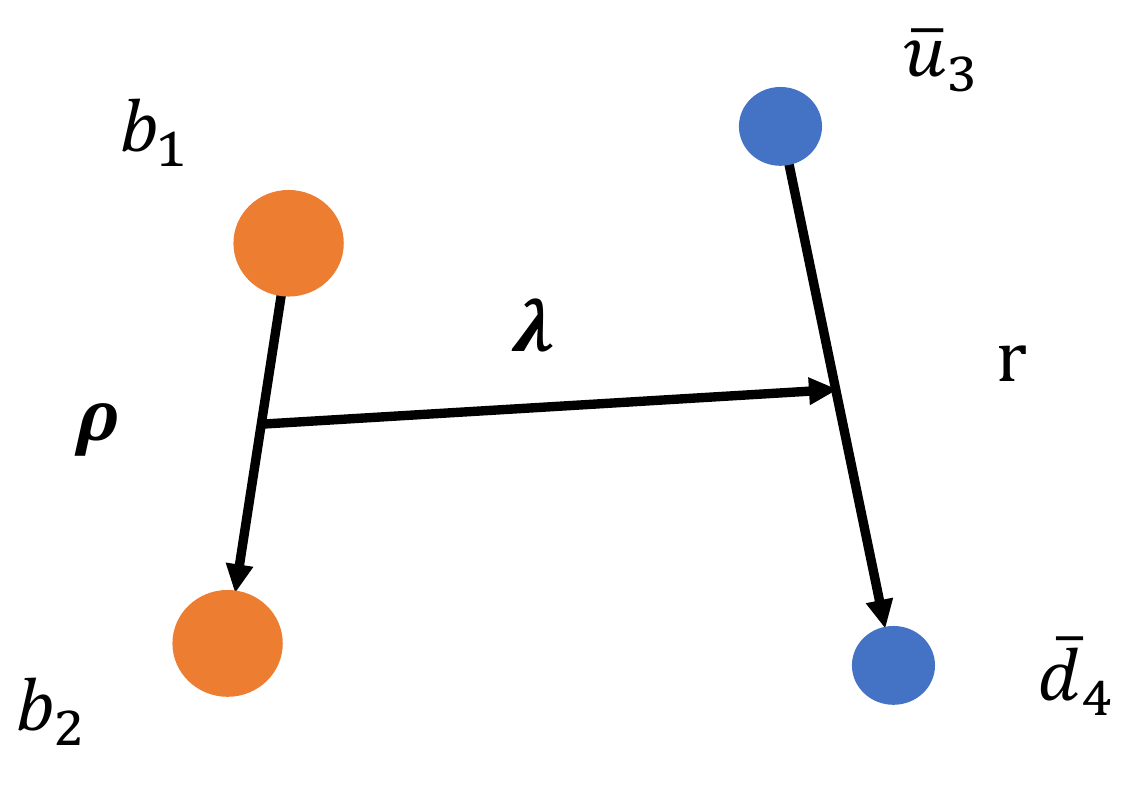}}
\caption{Jacobi coordinates of the states $T^-_{bb}$. Orange balls stand for $b$-quarks
and blue balls represent $u$- and $d$-quark.}
\label{jaccobi}
\end{figure}

In the diquark configuration, the Jacobi coordinates of the states $T^-_{bb}$ are
presented in Fig.~\ref{jaccobi}.
Their specific expressions can be written as
\begin{eqnarray}
\begin{aligned}
&\boldsymbol{\rho}=\mathbf{r}_1-\mathbf{r}_2,~~\mathbf{r}=\mathbf{r}_3-\mathbf{r}_4,\\
\noalign{\smallskip}
&\boldsymbol{\lambda}=\frac{m_1\mathbf{r}_1+m_2\mathbf{r}_2}{m_1+m_2}
-\frac{m_3\mathbf{r}_3+m_3\mathbf{r}_4}{m_3+m_4},
\end{aligned}
\end{eqnarray}
where $\mathbf{r}_i$ stand for the position of the $i$-th particle in the states
$T^-_{bb}$. Note that the coordinates are just a possible choice of many possibilities,
which is most propitious to describe the natures of diquarks and their correlations.
The orbital angular momentum associated with those coordinates is denoted as $l_{{\rho} }$,
$l_{r}$, and $l_{{\lambda}}$, respectively. In the present work, we mainly concentrate
on the $P$-wave excited states $T^-_{bb}$. The single $P$-wave excitation can take place
in the diquark $[bb]^{s_1}_{c_1}$ ($\rho $-mode: $l_{{\rho} }=1$), antidiquark
$[\bar{u}\bar{d}]^{s_2}_{c_2}$ ($r$-mode: $l_{r}=1$), or between the diquark $[bb]^{s_1}_{c_1}$
and antidiquark $[\bar{u}\bar{d}]^{s_2}_{c_2}$ ($\lambda$-mode: $l_{{\lambda}}=1$).
Similar orbital excited modes were adopted to study the doubly heavy tetraquark
states~\cite{Kim:2022mpa}.

Accurate numerical calculations are a primary requirement to exactly comprehend the
natures of the states $T^-_{bb}$. The Gaussian expansion method (GEM) has been proven
to be rather powerful to solve the few-body problem in nuclear physics~\cite{Hiyama:2003cu}.
A recent comparative study revealed the superiority of the GEM over the resonating
group method and the diffusion Monte Carlo method for the tetraquark bound states in
quark models~\cite{Meng:2023jqk}. According to the GEM, the relative motion wave
functions $\phi_{lm}(\mathbf{x})$ can be expanded as the superpositions of a set
of Gaussian functions with different sizes,
\begin{eqnarray}
\phi_{lm}(\mathbf{x})&=&\sum_{n=1}^{n_{max}}c_nN_{nl}x^{l}
e^{-\nu_nx^2}Y_{lm}(\hat{\mathbf{x}}),
\end{eqnarray}
where $\mathbf{x}$ represents $\boldsymbol{\rho} $, $\mathbf{r}$ and
$\boldsymbol{\lambda}$. More details about the GEM can be found in Ref.~\cite{Hiyama:2003cu}.

The color-spin configurations of the states $T^-_{bb}$ can be denoted as
$[bb]^{s_1}_{c_1}[\bar{u}\bar{d}]^{s_2}_{c_2}$, where the subscripts $c_i$ and
superscripts $s_i$ are the color and spin, respectively. Both the diquark
$[bb]^{s_1}_{c_1}$ and antidiquark $[\bar{u}\bar{d}]^{s_2}_{c_2}$ can be in the
spin singlet or triplet. The diquark $[bb]^{s_1}_{c_1}$ is an isospin singlet and
the antidiquark $[\bar{u}\bar{d}]^{s_2}_{c_2}$ can be in the isospin singlet or
triplet similar to its spin. The color representations of the diquark $[bb]^{s_1}_{c_1}$
can be antisymmetric $\bar{\mathbf{3}}$ and symmetric $\mathbf{6}$. Those of the
antidiquark $[\bar{u}\bar{d}]^{s_2}_{c_2}$ can be antisymmetric $\mathbf{3}$ and
symmetric $\bar{\mathbf{6}} $. The color configuration of the tetraquark state
$[bb][\bar{u}\bar{d}]$ can be written as
\begin{gather}
\begin{split}
(\bar{\mathbf{3}}\oplus\mathbf{6})\otimes(\mathbf{3} \oplus\bar{\mathbf{6}})
=\underbrace{(\bar{\mathbf{3}}\otimes\mathbf{3})}_{\mathbf{1}\oplus\mathbf{8}}
\oplus\underbrace{(\bar{\mathbf{3}} \otimes\bar{\mathbf{6}})}_{\mathbf{8}\oplus\overline{\mathbf{10}}}
\oplus\underbrace{(\mathbf{6}\otimes\mathbf{3})}_{\mathbf{8}\oplus\mathbf{10}}\oplus\underbrace{(\mathbf{6}
\otimes\bar{\mathbf{6}})}_{\mathbf{1}\oplus\mathbf{8}\oplus\mathbf{27}}\nonumber
\end{split}
\end{gather}
According to the colorless requirement, only two coupling modes,
$\bar{\mathbf{3}}\otimes\mathbf{3}$ and $\mathbf{6}\otimes\bar{\mathbf{6}}$, are permitted.
In general, the physical state should be the mixture of those two color coupling modes.

The diquark $[bb]^{s_1}_{c_1}$ and antidiquark $[\bar{u}\bar{d}]^{s_2}_{c_2}$ are a
spatially extended compound with various color-spin-isospin-orbit configurations.
Taking all degrees of freedom of identical particles into account, the Pauli principle
imposes some restrictions on the quantum numbers of the diquark $[bb]^{s_1}_{c_1}$ and
antidiquark $[\bar{u}\bar{d}]^{s_2}_{c_2}$~\cite{Deng:2022cld}. The total wave function
of the states $T^-_{bb}$ with the isospin $I$ and angular momentum $J$ can be constructed
as a sum of the following direct products of the diquark $[bb]^{s_1}_{c_1}$, antidiquark
$[\bar{u}\bar{d}]^{s_2}_{c_2}$, and their relative motion wave function
$\phi_{l_{{\lambda}}m_{{\lambda}}}(\boldsymbol{\lambda})$,
\begin{equation}
\begin{split}
\Phi^{T^-_{bb}}_{IJ}=\sum_{\alpha}c_{\alpha}
\left[\Psi^{[bb]}_{i_{1}j_{1}c_{1}l_{\rho }}\Psi^{[\bar{u}\bar{d}]}_{i_{2}j_{2}c_{2}l_{r}}
\phi_{l_{{\lambda}}m_{{\lambda}}}(\boldsymbol{\lambda})\right]^{T^-_{bb}}_{IJ}.~\label{wavefunction}
\end{split}
\end{equation}
The summation index $\alpha$ represents all of the possible color-spin-isospin-orbit
combinations that can be coupled into the total spin and isospin and the coefficients
$c_{\alpha}$ are determined by the model dynamics.

\section{numerical results and analysis}

We mainly concentrate on the natures of the $P$-wave excited states $T^-_{bb}$
in the isospin singlet and three excited modes, which are listed in Tables~\ref{diquark}
and \ref{tbb}.
\begin{table*}[ht]
\caption{Various energy distribution in the diquark $[bb]^{s_1}_{c_1}$, antidiquark
$[\bar{u}\bar{d}]^{s_2}_{c_2}$ and between them ($[bb]^{s_1}_{c_1}$-$[\bar{u}\bar{d}]^{s_2}_{c_2}$)
in the states $T^-_{bb}$, unit in MeV. The superscripts $\rho $, $r$
and $\lambda$ denote that the $P$-wave excitation occurs in the $[bb]^{s_1}_{c_1}$,
$[\bar{u}\bar{d}]^{s_2}_{c_2}$ and between them, respectively. $T$, $V^{\rm con}$,
$V^{\rm coul}$, $V^{\rm cm}$, $V^{\sigma}$, $V^{\pi}$, and $V^{\eta}$ are the kinetic
energy, confinement, Coulomb, chromomagnetic, $\sigma$-, $\pi$-, and $\eta$-meson-exchange,
respectively.} \label{diquark}
\tabcolsep=0.12cm
\begin{tabular}{cccccccccccccccccc}
\toprule[0.8pt]
\noalign{\smallskip}
\noalign{\smallskip}
$n^{2S+1}L_J$&Parts&~~~$T$~~~&$V^{\rm con}$&$V^{\rm cm}$&$V^{\rm coul}$&~~$V^{\eta}$~~&~$V^{\pi}$~&$V^{\sigma}$&
              Parts&~~~$T$~~~&$V^{\rm con}$&$V^{\rm cm}$&$V^{\rm coul}$&~~$V^{\eta}$~~&~$V^{\pi}$~&$V^{\sigma}$\\
\noalign{\smallskip}
\toprule[0.8pt]
\noalign{\smallskip}
\multirow{4}{*}{$1^{3}S_1$}           & $[bb]^1_{\bar{\mathbf{3}}}$                 &  124  &   16   &   1    & $-$199 &  0   &   0    &    0
                                      & $[bb]^0_{\mathbf{6}}$                       &  51   & $-$19  &   1    &   59   &  0   &   0    &    0  \\
\noalign{\smallskip}                  & $[\bar{u}\bar{d}]^0_{\mathbf{3}}$           &  789  &   55   & $-$289 & $-$257 &  57  & $-$335 & $-$40
                                      & $[\bar{u}\bar{d}]^1_{\bar{\mathbf{6}}}$     &  249  & $-$70  & $-$10  &   68   &  $-$2&   20   & $-$14 \\
\noalign{\smallskip}& $[bb]^1_{\bar{\mathbf{3}}}$-$[\bar{u}\bar{d}]^0_{\mathbf{3}}$ &  210  &  124   & $-$2   & $-$340 &  0   &   0    &    0
                    & $[bb]^0_{\mathbf{6}}$-$[\bar{u}\bar{d}]^1_{\bar{\mathbf{6}}}$ &  303  &  372   &    0   & $-$764 &  0   &   0    &    0  \\
\noalign{\smallskip}
\toprule[0.8pt]
\noalign{\smallskip}
\multirow{4}{*}{$1^{3}P_{0,1,2}^{\lambda}$} & $[bb]^1_{\bar{\mathbf{3}}}$           &  115  &   18   &   1    & $-$193 &  0   &   0    &    0
                                      & $[bb]^0_{\mathbf{6}}$                       &  41   &  $-$24 &   1    &   52   &  0   &   0    &    0  \\
\noalign{\smallskip}                  & $[\bar{u}\bar{d}]^0_{\mathbf{3}}$           &  739  &   59   & $-$273 & $-$250 &  53  & $-$316 & $-$38
                                      & $[\bar{u}\bar{d}]^1_{\bar{\mathbf{6}}}$     &  203  & $-$85  &  $-$7  &   60   &  $-$1&  14    & $-$11 \\
\noalign{\smallskip}& $[bb]^1_{\bar{\mathbf{3}}}$-$[\bar{u}\bar{d}]^0_{\mathbf{3}}$ &  282  &   216  & $-$1   & $-$232 &  0   &   0    &    0
                    & $[bb]^0_{\mathbf{6}}$-$[\bar{u}\bar{d}]^1_{\bar{\mathbf{6}}}$ &  449  &   552  &   0    & $-$604 &  0   &   0    &    0  \\
\noalign{\smallskip}
\toprule[0.8pt]
\noalign{\smallskip}
\multirow{4}{*}{$1^{1}P_1^{\rho}$}    & $[bb]^0_{\bar{\mathbf{3}}}$                 &  138  &   39   &    0   & $-$107 &  0   &   0    &    0
                                      & $[bb]^1_{\mathbf{6}}$                       &  97   & $-$28  &  $-$7  &   44   &  0   &   0    &    0  \\
\noalign{\smallskip}                  & $[\bar{u}\bar{d}]^0_{\mathbf{3}}$           &  775  &   55   & $-$283 & $-$253 &  56  & $-$328 & $-$40
                                      & $[\bar{u}\bar{d}]^1_{\bar{\mathbf{6}}}$     &  248  & $-$70  & $-$10  &   68   &  $-$2&   20   & $-$14 \\
\noalign{\smallskip}& $[bb]^0_{\bar{\mathbf{3}}}$-$[\bar{u}\bar{d}]^0_{\mathbf{3}}$ &  200  &  140   & $-$4   & $-$320 &  0   &   0    &    0
                    & $[bb]^1_{\mathbf{6}}$-$[\bar{u}\bar{d}]^1_{\bar{\mathbf{6}}}$ &  302  &  396   & $-$24  & $-$740 &  0   &   0    &    0  \\
\noalign{\smallskip}
\toprule[0.8pt]
\noalign{\smallskip}
\multirow{4}{*}{$1^{1}P_1^{r}$}       & $[bb]^1_{\bar{\mathbf{3}}}$                 &  116  &   17   &    1   & $-$195 &  0   &   0    &    0
                                      & $[bb]^0_{\mathbf{6}}$                       &   41  & $-$24  &    0   &   50   &  0   &   0    &    0  \\
\noalign{\smallskip}                  & $[\bar{u}\bar{d}]^1_{\mathbf{3}}$           &  435  &  218   &    4   & $-$93  &  1   &   2    &  $-$5
                                      & $[\bar{u}\bar{d}]^0_{\bar{\mathbf{6}}}$     &  375  & $-$125 &    4   &   42   & $-$1 &  $-$5  &  $-$4 \\
\noalign{\smallskip}& $[bb]^1_{\bar{\mathbf{3}}}$-$[\bar{u}\bar{d}]^1_{\mathbf{3}}$ &  165  &  228   &  $-$6  & $-$240 &  0   &   0    &    0
                    & $[bb]^0_{\mathbf{6}}$-$[\bar{u}\bar{d}]^0_{\bar{\mathbf{6}}}$ &  257  &  552   &  $-$2  & $-$596 &  0   &   0    &    0  \\
\noalign{\smallskip}
\toprule[0.8pt]
\end{tabular}
\caption{Binding energy $\Delta E$ and the contribution from various interactions and kinetic
energy to $\Delta E$, unit in MeV. $\langle\boldsymbol{\rho} ^2\rangle^{\frac{1}{2}}$ and
$\langle\mathbf{r}^2\rangle^{\frac{1}{2}}$ are the size of the diquark $[bb]^{s_1}_{c_1}$ and
antidiquark $[\bar{u}\bar{d}]^{s_2}_{c_2}$, respectively, and $\langle\boldsymbol{\lambda}^2\rangle^{\frac{1}{2}}$ is
their distance, unit in fm.} \label{tbb}
\tabcolsep=0.205cm
\begin{tabular}{cccccccccccccccccc}
\toprule[0.8pt]
\noalign{\smallskip}
\noalign{\smallskip}
$n^{2S+1}L_J$&Color-spin,~ratio&$\Delta E$&$\Delta T $&$\Delta V^{\rm con}$&
$\Delta V^{\rm cm}$&$\Delta V^{\rm coul}$&$\Delta V^{\eta}$&$\Delta V^{\pi}$
&$\Delta V^{\sigma}$&$\langle\boldsymbol{\rho} ^2\rangle^{\frac{1}{2}}$
&$\langle\mathbf{r}^2\rangle^{\frac{1}{2}}$
&$\langle\boldsymbol{\lambda}^2\rangle^{\frac{1}{2}}$\\
\noalign{\smallskip}
\toprule[0.8pt]
\noalign{\smallskip}
\multirow{3}{*}{$1^{3}S_1$}& $[bb]^1_{\bar{\mathbf{3}}}[\bar{u}\bar{d}]^0_{\mathbf{3} }$,~$>$99\%
& $-$215 &   479  &$-$50&$-$259 &  $-$67   &  57  &$-$335&  $-$40  &0.39&0.71&0.64\\
\noalign{\smallskip}
& $[bb]^0_{\mathbf{6}}[\bar{u}\bar{d}]^1_{\bar{\mathbf{6}}}$,~$<$1\%
&   119   &  $-$38  &  40  &  94   &    19    & $-$2 &  20  &  $-$14  &0.60&1.13&0.53\\
\noalign{\smallskip}
&Mixing
&   $-$216        &   481   &$-$51 &$-$260 &  $-$68  &  57  &$-$335 &  $-$40  &0.39&0.71&0.64\\
\noalign{\smallskip}
\toprule[0.8pt]
\noalign{\smallskip}
\multirow{3}{*}{$1^3P_{0,1,2}^{\lambda}$}& $[bb]^1_{\bar{\mathbf{3}}}[\bar{u}\bar{d}]^0_{\mathbf{3} }$,~$>99\%$
& 55  &  494  &  49  &$-$243&    56  &   53  &$-$316& $-$38 &0.40&0.74&0.91\\
\noalign{\smallskip}
& $[bb]^0_{\mathbf{6}}[\bar{u}\bar{d}]^1_{\bar{\mathbf{6}}}$,~$<1\%$
& 511  &   51  &  199 &  22  &   237  &  $-$1 &  14  & $-$11 &0.67&1.25&0.72\\
\noalign{\smallskip}
&Mixing
&       55     &  494  &  49  &$-$243&   56   &  53   &$-$316& $-$38 &0.40&0.74&0.90\\
\noalign{\smallskip}
\toprule[0.8pt]
\noalign{\smallskip}
\multirow{3}{*}{$1^{1}P_1^{\rho }$}& $[bb]^0_{\bar{\mathbf{3}}}[\bar{u}\bar{d}]^0_{\mathbf{3} }$,~$>$99\%
&  $-$18 &  409  &  7  &$-$210 &    91    &  56  &  $-$331  &  $-$40  &0.60&0.72&0.65\\
\noalign{\smallskip}
& $[bb]^1_{\mathbf{6} }[\bar{u}\bar{d}]^1_{\bar{\mathbf{6}}}$,~$<$1\%
&   203  & $-$53 &  69 &  39   &   144    & $-$2 &    20    &  $-$14  &0.72&1.13&0.53\\
\noalign{\smallskip}
&Mixing &     $-$18      &  412  &  6  &$-$213 &    89    &  56  &   $-$328 &  $-$40  &0.60&0.72&0.65\\
\noalign{\smallskip}
\toprule[0.8pt]
\noalign{\smallskip}
\multirow{3}{*}{$1^{1}P_1^{r}$}& $[bb]^1_{\bar{\mathbf{3}}}[\bar{u}\bar{d}]^1_{\mathbf{3} }$,~$<1\%$
&   562  &   15  & 233 &   74   &  243   &  1   &   2  &  $-$5  &0.40&1.42&0.71\\
\noalign{\smallskip}
& $[bb]^0_{\mathbf{6} }[\bar{u}\bar{d}]^0_{\bar{\mathbf{6}}}$,~$>99\%$
&   480& $-$30 & 176 &  80   &   265  & $-$1 & $-$5 &  $-$4 &0.67&1.53&0.57\\
\noalign{\smallskip}
&Mixing &   480    & $-$28 & 176 &  78   &  265   & $-$1 & $-$5 &  $-$4  &0.67&1.53&0.57\\
\noalign{\smallskip}
\toprule[0.8pt]
\end{tabular}
\end{table*}
In order to exhibit the influences of the $P$-wave excitation on the
states, we also present the natures of the ground state $T^-_{bb}$ with $1^3S_1$.
By accurately solving the four-body Schr\"{o}dinger equation with the well-defined
trial wave function, we can arrive at their eigenenergy and eigenwave function.
Using the eigenwave function, we can calculate various energy distributions in
the diquark $[bb]^{s_1}_{c_1}$, antidiquark $[\bar{u}\bar{d}]^{s_2}_{c_2}$ and
between them in the states $T^-_{bb}$, which are listed in Table~\ref{diquark}.
Subsequently, we can achieve their binding energy, $\Delta E=E_4-M_{b\bar{u}}-M_{b\bar{d}}$,
and the contributions coming from each part of the Hamiltonian to $\Delta E$ to show
the underlying dynamic mechanism in detail, where $E_4$ is the minimal eigenenergy
and $M_{b\bar{u}}+M_{b\bar{d}}$ is its lowest two-meson threshold. In addition, we
calculate the average distances and the ratio of each color spin
$[bb]^{s_1}_{c_1}[\bar{u}\bar{d}]^{s_2}_{c_2}$ in the coupled channels. We present
those numerical results in Table~\ref{tbb}.

\subsection{Ground state $T^-_{bb}$ with $1^3S_1$}

The ground state $T^-_{bb}$ with $1^3S_1$ is composed of two possible color-spin
configurations $[bb]^1_{\bar{\mathbf{3}}}[\bar{u}\bar{d}]^0_{\mathbf{3}}$
and $[bb]^0_{\mathbf{6} }[\bar{u}\bar{d}]^1_{\bar{\mathbf{6}}}$. The chromomagnetic
interaction and meson-exchange interactions in the antidiquark $[\bar{u}\bar{d}]^0_{\mathbf{3} }$
can provide extremely strong attractions about 600 MeV, see Table~\ref{diquark}.
The Coulomb interaction in the diquark $[bb]^1_{\mathbf{3} }$ and antidiquark
$[\bar{u}\bar{d}]^0_{\mathbf{3}}$ also gives strong attractions. These strong
attractions are beneficial to establish the deeply compact bound state $T^-_{bb}$
with $1^3S_1$ because they do not appear in the threshold $\bar{B}\bar{B}^*$.
In strong contrast, those attractions in the diquark $[bb]^0_{\mathbf{6}}$
and antidiquark $[\bar{u}\bar{d}]^1_{\bar{\mathbf{6}}}$ are very weak even repulsive, 
see Table~\ref{diquark}. Moreover, the color-electric interaction, i.e. confinement 
potential plus Coulomb interaction, between the diquark $[bb]^0_{\mathbf{6} }$ and antidiquark
$[\bar{u}\bar{d}]^1_{\bar{\mathbf{6}}}$ is much stronger than that between the
diquark $[bb]^1_{\bar{\mathbf{3}}}$ and antidiquark $[\bar{u}\bar{d}]^0_{\mathbf{3}}$.
The color-electric interaction between two colored subclusters in the color
configurations $\bar{\mathbf{3}}\otimes\mathbf{3}$ and $\mathbf{6}\otimes\bar{\mathbf{6}}$
has been discussed in detail in Refs.~\cite{Deng:2020iqw,Wang:2023vtx}. On the
whole, the mass of the configuration $[bb]^1_{\bar{\mathbf{3}}}[\bar{u}\bar{d}]^0_{\mathbf{3}}$
is much lower, about 334 MeV, than that of the configuration
$[bb]^0_{\mathbf{6}}[\bar{u}\bar{d}]^1_{\bar{\mathbf{6}}}$. In fact, the relative
strength of those interactions is also revealed by the average distances in
Table~\ref{tbb}. The stronger those interactions, the shorter the distances.

The color-spin configuration $[bb]^1_{\bar{\mathbf{3}}}[\bar{u}\bar{d}]^0_{\mathbf{3}}$
has a binding energy around $-215$ MeV in comparison to the threshold $\bar{B}\bar{B}^*$,
see Table~\ref{tbb}. In principle, the ground state $T^-_{bb}$ should be the mixture
of the two color-spin configurations. After their channel coupling calculation, the
binding energy of the state reduces to $-216$ MeV so that the state is a deeply compact
bound state, which is in good agreement with the conclusions of recent lattice
calculations~\cite{Francis:2016hui,Junnarkar:2018twb,Mohanta:2020eed,Leskovec:2019ioa}.
The color-spin configuration $[bb]^1_{\bar{\mathbf{3}}}[\bar{u}\bar{d}]^0_{\mathbf{3} }$
absolutely predominates the properties of the ground state while the color-spin configuration
$[bb]^0_{\mathbf{6} }[\bar{u}\bar{d}]^1_{\bar{\mathbf{6}}}$ can be completely ignored,
which is completely consistent with the conclusion in Ref.~\cite{Lu:2020rog}.
The vast majority of the binding energy comes from the chromomagnetic interaction and
meson-exchange interactions in the antidiquark $[\bar{u}\bar{d}]^0_{\mathbf{3}}$.
In addition, the color-electric interaction also contributes an attraction about
120 MeV to the binding energy. The kinetic energy provides a strong repulsion, which
is an obvious obstacle to forming this bound state.

In our previous work, see case (b) in Ref.~\cite{Deng:2021gnb}, we obtained a relative
loose molecular state $T^-_{bb}$ with $01^+$ using the same model Hamiltonian and parameters,
where the binding mechanism is the weak residual interaction between two colorless mesons.
Combining with the present work, the state $T^-_{bb}$ with $01^+$ can simultaneously exist
in two different structures in the same model, which also takes place in the similar model
study of the state~\cite{Meng:2020knc}. The phenomenon was deemed as a hadronic analog of
cluster formation in spectra of light nuclei, where cluster structures made of $\alpha$
particles are developed around the $\alpha$ emission thresholds, while the lower bound
states are compact shell-model-like states~\cite{Meng:2020knc}. We discussed the correlation
between two structures based on the angular momentum algebra and thought that their difference
comes from the different model spaces induced by different orbit excited modes~\cite{Deng:2022cld}.

\subsection{P-wave states $T^-_{bb}$ with $1^3P_{0,1,2}^{\lambda}$}.
In the excited states, the orbital excitation occurs between the
diquark $[bb]^{s_1}_{c_1}$ and antidiquark $[\bar{u}\bar{d}]^{s_2}_{c_2}$ so that their color-spin
configurations are exactly the same with those of the ground state $T^-_{bb}$ with $^3S_1$,
see Table~\ref{diquark}. In this calculation, we do not introduce spin-orbit interaction
and thus the states $1^3P_{0,1,2}^{\lambda}$ are degenerate. We expect that mass differences
among the states will be tiny since the differences are suppressed by the heavy quarks.
The natures of the diquark $[bb]^{s_1}_{c_1}$ and antidiquark $[\bar{u}\bar{d}]^{s_2}_{c_2}$
do not obviously change in the excited state $T^-_{bb}$ in comparison to the ground state
$T^-_{bb}$. However, the color-electric interactions between the diquark $[bb]^{s_1}_{c_1}$
and antidiquark $[\bar{u}\bar{d}]^{s_2}_{c_2}$ increase significantly, which remarkably
elevates the energy of the $P$-wave excited state $T^-_{bb}$ with $1^3P_{0,1,2}^{\lambda}$
relative to the ground state.

The mass of the color-spin configuration $[bb]^1_{\bar{\mathbf{3}}}[\bar{u}\bar{d}]^0_{\mathbf{3}}$
is much lower by about 450 MeV than that of the color-spin configuration
$[bb]^0_{\mathbf{6}}[\bar{u}\bar{d}]^1_{\bar{\mathbf{6}}}$ because of their
extremely different meson-exchange and chromomagnetic interactions between the
antidiquarks $[\bar{u}\bar{d}]^0_{\mathbf{3}}$ and $[\bar{u}\bar{d}]^1_{\bar{\mathbf{6}}}$.
The channel coupling calculation indicates that the color-spin configuration
$[bb]^1_{\bar{\mathbf{3}}}[\bar{u}\bar{d}]^0_{\mathbf{3}}$ completely dominates
the natures of the excited states $T^-_{bb}$ with $1^3P_{0,1,2}^{\lambda}$.
Their masses are 55 MeV higher than its threshold $\bar{B}\bar{B}^*$, the main
reason being that the $P$-wave excitation between the diquark $[bb]^1_{\bar{\mathbf{3}}}$
and the antidiquark $[\bar{u}\bar{d}]^0_{\mathbf{3}}$ enhances the
color-electric interaction.

\subsection{P-wave state $T^-_{bb} $ with $1^1P_1^{\rho}$} 

In the excited state, the orbital excitation takes place in the diquark $[bb]^{s_1}_{c_1}$.
The state consists of two color-spin configurations $[bb]^0_{\bar{\mathbf{3}}}[\bar{u}\bar{d}]^0_{\mathbf{3}}$
and $[bb]^1_{\mathbf{6}}[\bar{u}\bar{d}]^1_{\bar{\mathbf{6}}}$. The properties
of the antidiquark $[\bar{u}\bar{d}]^{s_2}_{c_2}$ in the state are almost exactly
consistent with those in the ground state $T^-_{bb}$ and the correlation between
the diquark $[bb]^{s_1}_{c_1}$ and antidiquark $[\bar{u}\bar{d}]^{s_2}_{c_2}$ just
changes a little bit, see Table~\ref{diquark}. The Coulomb interaction in the antidiquark
$[bb]^0_{\bar{\mathbf{3}}}$ is reduced by about 92 MeV because the interaction is sensitive
to the distance change induced by the $P$-wave excitation. Other natures of the diquark
$[bb]^{s_1}_{c_1}$ do not dramatically vary although its orbit is in the $P$-wave excitation
because of the suppression of the large mass of $b$-quarks.

The dominant color-spin configuration of the state $T^-_{bb}$ with $1^1P_1^{\rho }$
is $[bb]^0_{\bar{\mathbf{3}}}[\bar{u}\bar{d}]^0_{\mathbf{3}}$, in which the
chromomagnetic and meson-exchange interactions in the antidiquark
$[\bar{u}\bar{d}]^0_{\mathbf{3}}$ can still provide an extremely strong attraction.
Relative to the threshold $\bar{B}\bar{B}$, the binding energy of the state is
about $-18$ MeV so that it can establish a compact bound state. Its binding mechanism
originates from the chromomagnetic and meson-exchange interactions in the antidiquark
$[\bar{u}\bar{d}]^0_{\mathbf{3}}$, see Table~\ref{tbb}. The color-electric interaction,
especially for the Coulomb interaction, is not a binding mechanism anymore because of
the $P$-wave excitation in the diquark $[bb]^0_{\bar{\mathbf{3}}}$. The contributions
from the Coulomb interaction in the excited state $T^-_{bb}$ with $1^1P_1^{\rho }$
and the ground state $T^-_{bb}$ with $1^3S_1$ are the main reason resulting in their
binding energy difference.

Bicudo {\it et al.} studied the state $T_{bb}^-$ with $01^-$, where the $P$-wave excitation
occurs in the diquark $[bb]$, using the lattice QCD potentials, Born-Oppenheimer
approximation and emergent wave method~\cite{Bicudo:2017szl}. Its mass is $10576^{-4}_{+4}$
MeV, which is close to our prediction on the state but 16 MeV higher than the threshold
$\bar{B}\bar{B}$. The state can decay into two $\bar{B}$ mesons via strong interaction
so that it is a resonance. Subsequently, Hoffmann {\it et al.} refined the investigation of
the state by including heavy quark spin effects via the mass difference between $B$ and
$B^*$ mesons~\cite{Hoffmann:2022jdx}. They did not find any indication for the existence
of the resonance~\cite{Hoffmann:2022jdx}.

\subsection{P-wave state $T^-_{bb} $ with $1^1P_1^{r}$} 

In the excited state, the orbital excitation occurs in the antidiquark
$[\bar{u}\bar{d}]^{s_2}_{c_2}$, which induces a huge impact on its natures relative
to its ground state because of the changes of its spin and size, especially for the
diquark $[\bar{u}\bar{d}]^1_{\mathbf{3}}$, see Table~\ref{diquark}. The meson-exchange
and chromomagnetic interactions in the antidiquark $[\bar{u}\bar{d}]^1_{\mathbf{3}}$
are weak, just several MeVs, while they can provide strong attractions, around 600 MeV,
in the antidiquark $[\bar{u}\bar{d}]^0_{\mathbf{3}}$. In addition, the color-electric
interaction in the antidiquark $[\bar{u}\bar{d}]^1_{\mathbf{3}}$ is much higher, about
320 MeV, than that in the antidiquark $[\bar{u}\bar{d}]^0_{\mathbf{3}}$. Relatively speaking,
the natures of the antidiquark $[\bar{u}\bar{d}]^0_{\bar{\mathbf{6}}}$ do not dramatically
vary in comparison to its ground state $[\bar{u}\bar{d}]^1_{\bar{\mathbf{6}}}$.

The total mass of the antidiquark $[\bar{u}\bar{d}]^1_{\mathbf{3}}$ is about 280 MeV
higher than that of the antidiquark $[\bar{u}\bar{d}]^0_{\bar{\mathbf{6}}}$ mainly because
of the color-electric interaction. The fact directly results in that the dominant color-spin
configuration in the excited state is $[bb]^0_{\mathbf{6}}[\bar{u}\bar{d}]^0_{\bar{\mathbf{6}}}$
rather than $[bb]^1_{\bar{\mathbf{3}}}[\bar{u}\bar{d}]^1_{\mathbf{3}}$. Therefore, the
color configuration $\mathbf{6}\otimes\bar{\mathbf{6}}$ should not be unhesitatingly
discarded in the tetraquark excited states. The mass of the color-spin configuration
$[bb]^0_{\mathbf{6}}[\bar{u}\bar{d}]^0_{\bar{\mathbf{6}}}$ is much higher, about 480
MeV, than its threshold $\bar{B}\bar{B}$, see Table~\ref{tbb}. The chromomagnetic and
color-electric interactions cannot provide any attractions. Furthermore, the meson-exchange
interactions just give an attraction of 10 MeV. However, the kinetic energy contributes
an attraction of about 30 MeV to the binding energy mainly because of the spatial extension
of the diquark $[bb]^0_{\mathbf{6}}$ and antidiquark $[\bar{u}\bar{d}]^0_{\bar{\mathbf{6}}}$
due to the absence of strong binding forces in them. The channel coupling calculation of the
two color-spin configurations does not change the mass of the state.

\subsection{P-wave states $T^-_{bb} $ in other models} 

The $P$-wave states $T_{bb}^-$ in the three excited modes were investigated in the potential
chiral-diquark model~\cite{Kim:2022mpa}. The states $T_{bb}^-$ were described as a threebody
system composed of two heavy quarks and an antidiquark only in the color configuration
$\bar{\mathbf{3}}\otimes\mathbf{3}$. The $P$-wave excited states $T^-_{bb}$ with $0^-$, $1^-$
and $2^-$ in the excited $\lambda$-mode were investigated in a constituent quark model~\cite{Meng:2021yjr},
where the four-body problem is solved in a variational method. None of the $P$-wave bound
states $T_{bb}^-$ can be found because their predicted masses are much farther away form their
corresponding threshold~\cite{Kim:2022mpa,Meng:2021yjr}. The predicted masses for the $01^-$
states in the $\rho$- and $\lambda$-mode are higher over 100 MeV higher than those of the present
work because of the absence of meson exchange interactions providing strong attractions,
which is held true for the ground state $T_{bb}^-$ with $01^+$. The excited state with
$01^-$ in the $\xi_P$-mode is close to that in the $r$-mode of the present work because
the meson exchange interaction is very weak~\cite{Kim:2022mpa}.

The tetraquark states with diquark configuration include two color configurations
$\bar{\mathbf{3}}\otimes\mathbf{3}$ and $\mathbf{6}\otimes\bar{\mathbf{6}}$. The
configuration $\bar{\mathbf{3}}\otimes\mathbf{3}$ is usually preferred over the
configuration $\mathbf{6}\otimes\bar{\mathbf{6}}$ in the studies of tetraquark states
with diquark configuration~\cite{Barabanov:2020jvn}. In fact, the configuration
$\mathbf{6}\otimes\bar{\mathbf{6}}$ plays an important role in some systems, such
as the fully-heavy tetraquark states~\cite{Deng:2020iqw,Wang:2019rdo,Wu:2024euj}.
For the $S$-wave state $T^-_{bb}$ with diquark configurations, its main configuration
is widely regarded as $\bar{\mathbf{3}}\otimes\mathbf{3}$ in various theoretical
studies~\cite{Huang:2023jec,Chen:2022asf}, which is supported by the comparative
research~\cite{Deng:2022cld,Lu:2020rog}. For the $P$-wave states $T^-_{bb}$ in
the $r$-mode, the present comparative study indicates that the color configuration
$\mathbf{6}\otimes\bar{\mathbf{6}}$ instead of $\bar{\mathbf{3}}\otimes\mathbf{3}$
is its dominant component~\cite{Kim:2022mpa}. In this way, the color configuration
$\mathbf{6}\otimes\bar{\mathbf{6}}$ needs to be handled discreetly in the tetraquark
states.

\section{Summary}

We study the $P$-wave excited states $T^-_{bb}$ in the isospin singlet and three
excited modes from diquarks with the Gaussian expansion method in the quark model.
We decode the dynamical natures of the diquark $[bb]^{s_1}_{c_1}$, antidiquark
$[\bar{u}\bar{d}]^{s_2}_{c_2}$ and their correlations in the states $T^-_{bb}$ by
decomposing the interactions from various sources in the quark model. The ground state
antidiquark $[\bar{u}\bar{d}]^0_{\mathbf{3}}$ can provide extremely strong attractions
coming from meson-exchange, Coulomb, and chromomagnetic interactions. Those interactions 
in the antidiquark $[\bar{u}\bar{d}]^{s_2}_{c_2}$ with other quantum numbers are weak 
even repulsive. The Coulomb interaction predominates the natures of the diquark
$[bb]^{s_1}_{c_1}$, especially $[bb]^1_{\bar{\mathbf{3}}}$ and $[bb]^0_{\bar{\mathbf{3}}}$
($P$-wave), because the interaction is proportional to $\frac{1}{r}$ and the large
mass of $b$-quarks allows two $b$-quarks to be as close as possible. The correlations
between the diquark $[bb]^{s_1}_{c_1}$ and antidiquark $[\bar{u}\bar{d}]^{s_2}_{c_2}$
through the color-electric interaction only depend on the color representations of
the states $T^-_{bb}$. Either in the ground state or in the $P$-wave states, the
correlations in the color configuration $\mathbf{6}\otimes\bar{\mathbf{6}}$ are
stronger than those in the color configuration $\bar{\mathbf{3}}\otimes\mathbf{3}$.

The dominant color-spin configurations of the states $T^-_{bb}$ with $^3P^{\lambda}_{0,1,2}$
and $^1P_1^{\rho}$ are $[bb]^1_{\bar{\mathbf{3}}}[\bar{u}\bar{d}]^0_{\mathbf{3}}$
and $[bb]^0_{\bar{\mathbf{3}}}[\bar{u}\bar{d}]^0_{\mathbf{3}}$, respectively, which
are more than $99\%$. In strong contrast, those of the state $T^-_{bb}$ with $^1P^{r}_{1}$
is $[bb]^0_{\mathbf{6} }[\bar{u}\bar{d}]^0_{\bar{\mathbf{6}}}$ instead of
$[bb]^1_{\bar{\mathbf{3}}}[\bar{u}\bar{d}]^1_{\mathbf{3}}$, which indicates that
the color configuration $\mathbf{6}\otimes\bar{\mathbf{6}}$ needs to be handled discreetly
in the tetraquark states. The mass of the state with $1^1P^{\rho}_1$ is $18$ lower
MeV than the threshold $\bar{B}\bar{B}$ so that it can form a compact bound state.
The meson-exchange and chromomagnetic interactions in the antidiquark
$[\bar{u}\bar{d}]^0_{\mathbf{3}}$ are responsible for its binding mechanism.
The masses of the other two excited modes are higher than their respective threshold
so that they may be resonances.

The discovery of the state $T^+_{cc}$ has opened a new window for the doubly heavy
tetraquark states. The current study on the states $T^-_{bb}$ may be beneficial to
understand their underlying behaviors and nonperturbative QCD dynamics. We sincerely
expect more theoretical and experimental investigations to inspect the tetraquark
states from various perspectives in the near future.

\acknowledgments {This research is supported by the Chongqing Natural Science Foundation
under Project No. cstc2021jcyj-msxmX0078 and the Fundamental Research Funds for the
Central Universities under Grant No. SWU-XDJH202304.}

\end{document}